\documentclass[notitlepage,superscriptaddress,showpacs,showkeys, nofootinbib]{revtex4-1}

\usepackage{latexsym}
\usepackage{amsfonts}
\usepackage{amsbsy}
\usepackage{mathrsfs}
\usepackage[colorlinks=true,linkcolor=blue,citecolor=red,urlcolor=blue]{hyperref}

\newcommand{\rmi}{{\rm i}}
\newcommand{\JMP}{J. Math. Phys. }
\newcommand{\CQG}{Class. Quantum Grav. }
\newcommand{\PRL}{Phys. Rev. Lett. }
\newcommand{\PR}{Phys. Rev. }

\begin{document}

\title[Gauge and spacetime]{Gauge and spacetime connections in the Pleba\'nski\\ formulation of general relativity}

\author{Diego Gonz\'alez}
\email{dgonzalez@fis.cinvestav.mx}
\affiliation{Departamento de F\'{\i}sica, Cinvestav, Instituto Polit\'ecnico Nacional 2508,\\
San Pedro Zacatenco, 07360, Gustavo A. Madero, Ciudad de M\'exico, M\'exico}

\author{Merced Montesinos}
\email{merced@fis.cinvestav.mx}
\homepage{http://www.fis.cinvestav.mx/~merced/}
\affiliation{Departamento de F\'{\i}sica, Cinvestav, Instituto Polit\'ecnico Nacional 2508,\\
San Pedro Zacatenco, 07360, Gustavo A. Madero, Ciudad de M\'exico, M\'exico}

\author{Mercedes Vel\'azquez}
\email{mquesada@fis.cinvestav.mx}
\affiliation{Departamento de F\'{\i}sica, Cinvestav, Instituto Polit\'ecnico Nacional 2508,\\
San Pedro Zacatenco, 07360, Gustavo A. Madero, Ciudad de M\'exico, M\'exico}
\affiliation{Centre de Physique Th\'eorique, Campus de Luminy, 13288 Marseille, France}

\date{\today}

\begin{abstract}
The Pleba\'nski formulation of complex general relativity given in terms of variables valued in the complexification of the $so(3)$ Lie algebra is genuinely a gauge theory that is also diffeomorphism-invariant. For this reason, the way that the spin connection emerges from this formulation is not direct because both the internal (gauge) and the spin connections  are geometrical structures {\it a priori} not related; i.e., there is not a natural link between them. Their relationship must be put in by hand or must come from extra hypotheses. In this paper, we analyze the correct relationship between these connections and show how they are related. Our approach is different from the usual one in the sense that we do not assume a priori a spin connection from the very beginning, but employs the most general spacetime connection allowed in the first-order formalism.
\end{abstract}



\maketitle

\section{Introduction}
In any gauge theory based on internal (gauge) connections there are no concepts like spacetime connection, torsion, non-metricity, etc. {\it A priori}, internal connections have nothing to do with spacetime connections. Alternatively, the internal (gauge) bundle associated to any internal (gauge) theory (e.g. Pleba\'nski's formulation) has nothing to do with the tangent bundle associated to a spacetime connection employed in gravity theories (e.g. general relativity) \cite{trautman}. As a matter of illustration take, as an example, the Chern-Simons theory defined on a 3-dimensional manifold and given by the Lagrangian $L_{CS} = \left ( A^i \wedge d A^j + \frac{1}{3} f^i{}_{kl} A^k \wedge A^l \wedge A^j \right ) k_{ij} $, where $A=A^iJ_i$ is a connection 1-form,  $J_i$ are the generators of the Lie algebra and satisfy the usual commutation relations $[J_i,J_j]= f^k{}_{ij} J_k$, $f^k{}_{ij}$ are the structure constants of the Lie algebra, and $k_{ij}=-\frac{1}{2}f^k{}_{il}f^l{}_{jk}$ is the Killing-Cartan metric. There is no way to parallel transport any of the fields involved in this theory along a curve on the manifold with a spacetime  connection simply because the latter does not exist. Take now, the Yang-Mills theory defined on a 4-dimensional manifold and given by the Lagrangian $L_{YM}= \left ( F^i \wedge \ast F^j \right ) k_{ij}$, where $F^i= d A^i + \frac12 f^i{}_{jk} A^j \wedge A^k$ is the field strength and ${\ast  F}^i$ its Hodge dual. In this case, we have a metric tensor from which we compute the Hodge dual $\ast$. Moreover, we can introduce in this framework the Levi-Civita (spacetime) connection associated with the metric tensor \cite{civita}. Even if we do that, the Levi-Civita (spacetime) connection and the internal connection $A^a$ from which we compute the Yang-Mills field strength $F^i$ are unrelated objects.

The Pleba\'nski formulation of complex general relativity \cite{plebanski} is closer to the Chern-Simons theory  than to the Yang-Mills theory in the sense that all the fields involved are valued in the complexification of the $so(3)$ Lie algebra, and there is not a spacetime metric from the very beginning nor a spacetime connection in such a formulation. Therefore, the way Einstein's equations of motion for general relativity emerge from the Pleba\'nski equations of motion is something non-trivial,  non-direct, and non-clear. In particular, if the spin connection has to emerge somehow and additionally if it also has to be related in some way to the internal (gauge) connection of the Pleba\'nski formulation, some extra hypotheses have to be made. The reader might think that the link between the gauge and the spacetime connections is something provided by Pleba\'nski's equations of motion. Unfortunately this is not so and this point is usually overlooked. 

Pleba\'nski's formulation for general relativity was rediscovered by Capovilla {\it et al.} in Ref. \cite{capo}. However, the relationship between the gauge and the spacetime connections was left there at the same status as it is in Pleba\'nski's paper. Later, this issue was also analyzed in the same context by Bengtsson in Ref. \cite{bengtsson} or more recently in \cite{krasnov4,krasnov5}.  Nevertheless, in all these papers the link between the internal and the spacetime connections is done assuming a spin connection introduced in by hand. This is the usual approach reported in the literature.

In this paper we also study this topic but our approach is different from the usual one in the sense that we do {\it not} assume {\it a priori} a spin connection \cite{civita}. We give the {\it exact} relationship between the internal connection and the spacetime connection within the most general framework allowed by the first Cartan structural equation involving non-vanishing torsion \cite{cartan} and by a spacetime connection with non-vanishing non-metricity. Furthermore, our approach displays clearly all the building blocks used in the construction of the spacetime connection from the internal (gauge) connection. Surprisingly, at the end of the calculations the internal connection becomes again the self-dual part of the spin connection no matter which torsion and non-metricity is involved in the general spacetime connection used from the very beginning. The fact that it is not necessary to restrict the analysis to the torsionless nor vanishing non-metricity case is a non-trivial and unexpected result, and it is the subject of the present paper.

\section{Pleba\'nski formulation of complex general relativity}
Pleba\'nski equations of motion define a genuine gauge theory that is also diffeomorphism-invariant. The equations of motion for the theory are [see the appendix \ref{appendixA}]
\begin{eqnarray}
&& \delta \Sigma^i: F^i = C^i{}_j \Sigma^j, \quad C^i{}_i=0, \label{einstein}\\
&& \delta A^i:  d \Sigma^i + \varepsilon^i{}_{jk} A^j \wedge \Sigma^k=0, \label{pleb}\\
&& \delta C_{ij}: \Sigma^i \wedge \Sigma^j - \frac13 \delta^{ij} \Sigma^k \wedge \Sigma_k=0, \label{pleb2}
\end{eqnarray}
where all the fields involved are valued in the complexification of the $so(3)$ Lie algebra.  One of the most amazing properties of the Pleba\'nski formulation is that literally there is no spacetime metric, spacetime connection, vielbeins, torsion (vanishing or not), non-metricty (vanishing or not) nor any other geometrical structure usually found in metric theories for gravity. For this reason, the way that general relativity emerges from Eqs. (\ref{einstein}), (\ref{pleb}), and (\ref{pleb2}) is something non-direct.

The road towards the spacetime structures begins with Eq. (\ref{pleb2}), which has, modulo the reality conditions
\begin{eqnarray}
&&\Sigma^i \wedge {\overline \Sigma}^j = 0, \label{reality1}\\
&&\Sigma^i \wedge \Sigma_i + {\overline \Sigma}^i \wedge {\overline \Sigma}_i =0, \label{reality2}
\end{eqnarray}
the solutions
\begin{eqnarray}
&& \Sigma^i = \theta^0 \wedge \theta^i +  \frac{\rmi}{2} \varepsilon^i{}_{jk} \theta^j \wedge \theta^k, \label{sol}\\
&& {\overline\Sigma}^i = \theta^0 \wedge \theta^i -  \frac{\rmi}{2} \varepsilon^i{}_{jk} \theta^j \wedge \theta^k, \label{sol2}
\end{eqnarray}
in the non-degenerate case, where $\{ \theta^0 , \theta^i \}$ is a set composed of four linearly independent real 1-forms.

The second step consists in substituting any of the two solutions, $\Sigma^i$ or ${\overline \Sigma}^i$, into Eq. (\ref{pleb}) and to solve for the internal connection $A^i$.

In what follows we will look for the solution for the connection in the case of $\Sigma^i$ given by (\ref{sol}). In order to do that, we realize that (\ref{pleb}) is an inhomogeneous linear system of equations for the components contained in $A^i$. The solution of (\ref{pleb}) is, as is shown in the appendix \ref{appendixB} (compare also with Ref. \cite{bengtsson})
\begin{eqnarray}
A^i = - \frac{1}{3!} \Psi^i{}_{jTR} d \Sigma^j{}_{IJK} {\widetilde\eta}^{IJKR} \theta^T, \label{connex}
\end{eqnarray}
with
\begin{eqnarray}
\Psi^{ij}{}_{MN} = -8 \left( \frac{m^{ij} k_{MN}}{\det{(m^{ij})}} +\frac{1}{4} \varepsilon^{ijk} (m^{-1})_{kl} \Sigma^l{}_{MN} \right ), \label{connexPsi}
\end{eqnarray}
where
\begin{eqnarray}
{k}_{MN} := -\frac{1}{12} \varepsilon_{ijk} \Sigma^i{}_{MI} \Sigma^j{}_{JK} \Sigma^k{}_{LN} \ {\widetilde \eta}^{IJKL},  \label{urbmetric}
\end{eqnarray}
\begin{eqnarray}
m^{ij}:=\frac12 \Sigma^i{}_{IJ} \Sigma^j{}_{KL} \ {\widetilde \eta}^{IJKL}, \label{mpleb}
\end{eqnarray}
which holds for generic $\Sigma^i$ provided that $\det{(m^{ij})}\neq 0$ (all details are in the appendix \ref{appendixB}). Equation (\ref{connex}) is equivalent to Eq. (\ref{pleb}), it is just differently written. Furthermore, if $\Sigma^i$ is given by (\ref{sol}), equations (\ref{urbmetric}) and (\ref{mpleb}) become $(k_{IJ})=(\eta_{IJ})= \mbox{diag} (-1,+1,+1,+1)$ and $m^{ij}=4 \rmi \delta^{ij}$, and equations (\ref{connex}) and (\ref{connexPsi}) acquire the form \cite{thesisdiego,tdiego}
\begin{eqnarray}
A^i = - \frac{1}{3!} \Psi^i{}_{jTR} d \Sigma^j{}_{IJK} {\widetilde\eta}^{IJKR} \theta^T, \label{diego}
\end{eqnarray}
with
\begin{eqnarray}
\Psi^{ij}{}_{MN} = \frac12 \left ( \delta^{ij} \eta_{MN} + \rmi \, \varepsilon^{ij}{}_k \Sigma^k{}_{MN} \right ).
\end{eqnarray}
More clearly, for $\Sigma^i$ given by (\ref{sol}) the nature and the status of the connection $A^i$ in (\ref{connex}) and in (\ref{diego}) is the same, i.e., the connection given in Eq. (\ref{diego}) is {\it still} a gauge connection, {\it not} a spacetime connection! The gauge connection $A^i$ (\ref{connex}) did not become a spacetime connection just by using Eq. (\ref{sol}).

\section{From the internal connection to the spacetime connection}
In order to link the internal connection (\ref{diego}) with a spacetime  connection some additional assumptions have to be put in by hand. We emphasize the readers that this link {\it does not} follow from the Pleba\'nski equations of motion themselves (\ref{einstein}), (\ref{pleb}), and (\ref{pleb2}). This remark is important because  people usually take for granted that:

 ``The system of equations (\ref{pleb}) is linear and therefore has a unique solution, and due to the fact that
\begin{eqnarray}
A^i = \rmi  \omega^i{}_0 + \frac12 \varepsilon^{ij}{}_k \omega^k{}_j, \label{ashtekar}
\end{eqnarray}
satisfies (\ref{pleb}), this is the only solution''. The statement contained in the quotation  marks is part of the usual approach found in the literature.

Nevertheless, it is not correct to say at this stage that  Eq. (\ref{ashtekar}) is the solution of Eq. (\ref{pleb}) simply because Eq. (\ref{ashtekar}) involves a spin connection whose link with the internal connection has not been given yet, i.e., the only knowledge of $\Sigma^i$ is not enough for $A^i$ is related to a spacetime connection via Eq. (\ref{pleb}).

In opposition to an internal connection $A^i$, a 4-dimensional space-time connection $\Gamma^I{}_J$ is an object defined by the following equations of motion \cite{civita,cartan,torres}
\begin{eqnarray}
D \theta^I := d \theta^I + \Gamma^I{}_J \wedge \theta^J = T^I, \label{def} \\
D g_{IJ} := d g_{IJ} - \Gamma^K{}_I g_{KJ} - \Gamma^K{}_J g_{IK} = M_{IJ}, \label{def2}
\end{eqnarray}
where $T^I$ are the torsion 2-forms and $M_{IJ}=M_{IJK}\theta^K$ are the non-metricity 1-forms such that $M_{IJ}=M_{JI}$. Here, $\{\theta^1, \theta^2, \theta^3, \theta^4\}$ is a set of 1-forms that form the dual basis (that is, $\theta^I (e_J) = \delta^I_J$) to the basis $\{e_1, e_2, e_3, e_4\}$ of the tangent space at each point of the 4-dimensional spacetime, and $g_{IJ}$ are the components of a metric tensor $g$ with respect to the basis $\{e_I \}$, i.e., $g_{IJ} = g (e_I, e_J)$. The system of Eqs. (\ref{def}) and (\ref{def2}) is closed in the sense that the number of equations is the same as the number of variables contained in  $\Gamma^I{}_J$. Therefore, once $T^I$ and $M_{IJ}$ are given, the connection  is uniquely defined by these equations\footnote{Strictly speaking, the definition of a spacetime connection is more general than the definition given by Eqs. (\ref{def}) and (\ref{def2}) (see, for instance, Ref. \cite{torres}). From the viewpoint of this more general definition, Eqs. (\ref{def}) and (\ref{def2}) are just a way to define a connection. Nevertheles, this way of defining the connection is usually the one employed in metric and gravitational theories, and will be used also here.}. The curvature $R^I{}_J$ of the spacetime connection $\Gamma^I{}_J$ is given by the second Cartan structural equation $R^I{}_J = d \Gamma^I{}_J + \Gamma^I{}_K \wedge \Gamma^K{}_J$.

The particular connection, denoted by $\omega^I{}_J$, and defined by the equations $T^I=0$ and $M_{IJ}=0$ is called the Levi-Civita or spin connection \cite{civita}
\begin{eqnarray}
d \theta^I + \omega^I{}_J \wedge \theta^J = 0, \label{def3} \\
d g_{IJ} - \omega^K{}_I g_{KJ} - \omega^K{}_J g_{IK} = 0. \label{def4}
\end{eqnarray}

Once we have reminded the reader the basic notions involved in the definition of a spacetime connection, it is time to pose the problem addressed in this paper, namely to show the exact relationship between $\Gamma^I{}_J$, $\omega^I{}_J$, and $A^i$. There is a priori no reason to discard $\Gamma^I{}_J$ in the analysis and to restrict ourselves to  only use  $\omega^I{}_J$.  In order to set the link between $\Gamma^I{}_J$, $\omega^I{}_J$, and $A^i$, we need to use the hypotheses 1 and 2 contained in each of the two next subsections. The difference between the two subsections is the spacetime metric used in the hypothesis 1.

\subsection{Using the Urbantke metric defined by the $\Sigma$'s} 

In order to continue, we impose:

{\bf Hypothesis 1}. First of all, we must involve the spacetime connection $\Gamma^I{}_J$ defined in Eqs. (\ref{def}) and (\ref{def2}), but this definition involves a metric $g_{IJ}$. So, which metric $g$ should be chosen and why? We recall that the expression for the internal connection (\ref{connex}) naturally involves the Urbantke metric (\ref{urbmetric}) \cite{urbantke}, so it seems logic to use this metric in (\ref{def}) and (\ref{def2}), i.e., to take $g_{IJ}=k_{IJ}$ and therefore $g_{IJ}=\eta_{IJ}$ because $k_{IJ}=\eta_{IJ}$ when $\Sigma^i$ is given by (\ref{sol})\footnote{ In local coordinates the Urbantke metric (\ref{urbmetric}) reads $-\frac{1}{12} \varepsilon_{ijk} \Sigma^i{}_{\mu\alpha} \Sigma^j{}_{\beta\gamma} \Sigma^k{}_{\delta\nu} {\widetilde \eta}^{\alpha\beta\gamma\delta}$.  When the expression for  $\Sigma^i$ given in (\ref{sol}) is inserted into the right-hand side of this equation, it becomes $ \det{(\theta^I_{\mu})} \theta^I_{\mu} \theta^J_{\nu} \eta_{IJ}$, which amounts to say that the tetrads $\theta^I$ are orthonormal.}.

{\bf Hypothesis 2}. What about values of torsion $T^I$ and non-metricity $M_{IJ}$ involved in (\ref{def}) and (\ref{def2}) respectively? We leave them totally arbitrary, i.e., they can be freely chosen.

Therefore, hypothesis 1 and 2 mean that we are going to consider the spacetime connection $\Gamma^I{}_J$ defined by the triplet $\{\eta_{IJ}, T^I, M_{IJ} \}$ via Eqs. (\ref{def}) and (\ref{def2}). Using the hypothesis 1 and partially the hypothesis 2 through (\ref{def}), we compute $d \Sigma^i$ and get
\begin{eqnarray}\label{key}
d \Sigma^i &=& \varepsilon^i{}_{jk} \left ( - \frac12 \varepsilon^j{}_{mn} \tau^{mn} + \rmi  \tau^{0j} \right ) \wedge \Sigma^k - \frac12 \left ( \Gamma^0{}_0 + \Gamma_j{}^j \right ) \wedge \Sigma^i \nonumber\\
&& + \frac{\rmi}{2} \varepsilon^{ij}{}_k \left ( \Gamma_{0j} - \Gamma_{j0} \right ) \wedge \Sigma^k - \frac12 \left ( \Gamma^i{}_j - \Gamma_j{}^i\right ) \wedge \Sigma^j \nonumber\\
&& - \frac12 \left ( \Gamma^0{}_0 - \Gamma_j{}^j \right ) \wedge {\overline \Sigma}^i - \frac{\rmi}{2} \varepsilon^{ij}{}_k \left ( \Gamma_{0j} +\Gamma_{j0} \right ) \wedge {\overline \Sigma}^k \nonumber\\
&& - \frac12 \left ( \Gamma^i{}_j + \Gamma_j{}^i \right ) \wedge{\overline \Sigma}^j,
\end{eqnarray}
where we have rewritten the torsion 2-forms $T^I$ as $T^I = \tau^I{}_J \wedge \theta^J$ with $\tau^{IJ}= -\tau^{JI}$. Notice the contribution of the anti-self-dual 2-forms ${\overline \Sigma}^i$. 

It is important to recall that in the usual approach both torsion and non-metricity are set to be zero ($T^I=0$ and $M_{IJ}=0$) from the very beginning. Therefore, $\Gamma^I{}_J \rightarrow \omega^I{}_J$ according to (\ref{def3}) and (\ref{def4}) and thus Eq. (\ref{key})  becomes
\begin{eqnarray}\label{usual}
d \Sigma^i = - \varepsilon^i{}_{jk} \left (  \rmi  \omega^{0j} - \frac12 \varepsilon^j{}_{mn} \omega^{mn} \right ) \wedge \Sigma^k {} .
\end{eqnarray}
When (\ref{usual}) is inserted into Eq. (\ref{pleb}) or Eq. (\ref{diego}), one gets (\ref{ashtekar}), which is the usual approach reported in literature. 

However, we are not going to proceed along the lines of the previous paragraph. Why? Precisely because we want to explore the consequences of leaving $T^I$ and $M_{IJ}$ totally arbitrary in (\ref{def}) and (\ref{def2}), without restricting the analysis to the use of the spin connection or any other particular connection. We are going to use the generic expression given in Eq. (\ref{key}) and to substitute it into (\ref{pleb}) or, equivalently, into (\ref{diego}) to obtain
\begin{eqnarray}\label{conintprev}
A^i &=& \left ( \rmi  \Gamma^{0i} - \frac12 \varepsilon^i{}_{jk} \Gamma^{jk} \right ) - \left ( \rmi  \tau^{0i}  -
\frac12 \varepsilon^i{}_{jk}  \tau^{jk} \right ) \nonumber\\
&& + \frac12 \varepsilon^i{}_{jk} \left (  \Gamma ^j{}_L{}^k +  \Gamma{}_L{}^{jk} \right ) \theta^L
+ \frac{\rmi}{2} \left (  \Gamma^{i}{}_{0} +  \Gamma_{0}{}^{i} \right ) \nonumber\\
&& + \frac{\rmi}{2} \left(  \Gamma_{0L}{}^{i} + \Gamma_{L0}{}^{i} - \Gamma^{i}{}_{L0}- \Gamma_{L}{}^{i}{}_{0}  \right) \theta^L,
\end{eqnarray}
where $\Gamma^I{}_J=\Gamma^I{}_{JK} \theta^K$. Using Eq. (\ref{def2}) (and therefore the hypothesis 2) the last two lines of Eq. (\ref{conintprev}) are rewritten and the internal connection acquires the form
\begin{eqnarray}
A^i &=& \left ( \rmi  \Gamma^{0i} - \frac12 \varepsilon^i{}_{jk} \Gamma^{jk} \right ) - \left ( \rmi  \tau^{0i}  -
\frac12 \varepsilon^i{}_{jk}  \tau^{jk} \right ) \nonumber\\
&& - \frac12 \varepsilon^i{}_{jk} M^j{}_L{}^k \theta^L
+ \frac{\rmi}{2} M^{0i} + \frac{\rmi}{2} M^0{}_L{}^i \theta^L - \frac{\rmi}{2} M^i{}_L{}^0 \theta^L \nonumber\\
& = & \rmi \left ( \Gamma^{0i} - \tau^{0i} + \frac{1}{2} M^{0i} + \frac12 M^0{}_L{}^i \theta^L - \frac12 M^i{}_L{}^0 \theta^L \right ) \nonumber\\
&& - \frac12 \varepsilon^i{}_{jk} \left ( \Gamma^{jk} - \tau^{jk} + M^j{}_L{}^k \theta^L  \right ).\label{conint}
\end{eqnarray}
Last equation gives the general form that $A^i$ acquires once it is assumed that Eqs. (\ref{def}) and (\ref{def2}) hold
with arbitrary torsion, arbitrary non-metricity, and with the Minkowski metric involved in Eq. (\ref{def2}). At this
stage the link between internal connection and spacetime connection has been done through last assumptions. Once the
link is done, one can focus on rewriting Eq. (\ref{conint}) in a more convenient form. In fact, by using the general
solution of Eqs. (\ref{def}) and (\ref{def2}), which expresses the relation between $\Gamma^I{}_J$ and $\omega^I{}_J$
as
\begin{eqnarray}\label{conn}
\Gamma^I{}_J = \omega^I{}_J + \tau^I{}_J -  \frac12 M^I{}_{KJ} \theta^K + \frac12 M_{JK}{}^I \theta^K - \frac12 M^I{}_J,
\end{eqnarray}
it is simple to see that Eq. (\ref{conint}) reduces to Eq. (\ref{ashtekar})
\begin{eqnarray}
A^i = \rmi \omega^i{}_0 + \frac12 \varepsilon^{ij}{}_k \omega^k\,_j. \nonumber
\end{eqnarray}

\noindent What we have shown is that by using Eqs. (\ref{def}) and (\ref{def2}) for linking the internal  
connection $A^i$ given in Eq. (\ref{pleb}) to a spacetime connection, no matter which torsion $T^I$ and non-metricity $M_{IJ}$ we choose from the very beginning to define $\Gamma^I{}_J$ because the internal solution (\ref{diego}) will be expressed {\it only} in terms of the spin connection $\omega^{I}{}_{J}$, which is the first term in the formula for $\Gamma^I{}_J$ given in (\ref{conn}). This is a non-trivial result and is the main result of the paper. In other words, it is not necessary to restrict the analysis to the torsionless nor vanishing non-metricity case from the very beginning, all what is needed is to take into account properly the relations given in Eqs. (\ref{key}), (\ref{conintprev}), (\ref{conint}) y (\ref{conn}).

\subsection{Using the Urbantke metric defined by the $F$'s}

In SubSection 3.1, the spacetime metric was chosen to be the Urbantke metric $k_{MN}$ defined by the $\Sigma$'s via Eq. (\ref{urbmetric}), and torsion and non-metricity were left arbitrary. Alternatively, motivated by the fact that the metric introduced by Urbantke in Ref. \cite{urbantke} is given in terms of the curvature 2-forms by $-\frac{1}{12} \varepsilon_{ijk} F^i{}_{\mu\alpha} F^j{}_{\beta\gamma} F^k{}_{\delta\nu} {\widetilde \eta}^{\alpha\beta\gamma\delta}$ and that this, for the case of the Plebanski formulation, is conformally related to the metric defined by the $\Sigma$'s in the following way $\det{(C_{ij})} \left( -\frac{1}{12} \varepsilon_{ijk} \Sigma^i{}_{\mu\alpha} \Sigma^j{}_{\beta\gamma} \Sigma^k{}_{\delta\nu} {\widetilde \eta}^{\alpha\beta\gamma\delta}  \right) $ where $\det{(C_{ij})}$ is a function of the two invariants Tr$C^2$ and Tr$C^3$, it seems reasonable to generalize this fact and to take  
 \begin{eqnarray}
{h}_{MN} = f \, {k}_{MN},  \label{urbmetricF}
\end{eqnarray}
as the space-time metric where now $f$ is an arbitrary non-vanishing conformal factor. Note that $h_{MN}$ becomes $f \ {\eta}_{MN}$ when $\Sigma^i$ is given by (\ref{sol}). Therefore, the analog of hypotheses 1 and 2 is now:

{\bf Hypothesis 1}. The metric $g_{IJ}$ in (\ref{def2}) is given by $g_{IJ}= f\, \eta_{IJ}$.

{\bf Hypothesis 2}. Torsion $T^I$ and non-metricity $M_{IJ}$ are left arbitrary in (\ref{def}) and (\ref{def2}).

The strategy to link the internal connection $A^i$ with the spacetime connection $\Gamma^I{}_J$  ($\equiv \Omega^I{}_J$) of (\ref{def}) and (\ref{def2})  defined by the triplet $\{ h_{IJ}, T^I, M_{IJ} \}$ via Cartan's equations
\begin{eqnarray}
D \theta^I := d \theta^I + \Omega^I{}_J \wedge \theta^J = T^I, \label{defOmega} \\
D h_{IJ} := d h_{IJ} - \Omega^K{}_I h_{KJ} - \Omega^K{}_J h_{IK} = M_{IJ}, \label{defOmega2}
\end{eqnarray}
is completely analogous to the one presented in the SubSection 3.1. Then, following the same procedure, we found that the solution for the internal connection $A^i$ is
\begin{eqnarray}\label{Asol2}
A^i = \rmi \omega^{i}{}_{0} + \frac12 \varepsilon^{ij}{}_{k} \omega^{k}{}_{j},
\end{eqnarray}
because of
\begin{eqnarray}\label{connOmega}
\Omega^{I}{}_{J} &=& \omega^{I}{}_{J} + \tau^{I}{}_{J} -  \frac12 M^I{}_K{}_J \theta^K + \frac12 M_J{}_K{}^I \theta^K - \frac12 M^{I}{}_{J} \nonumber\\
&&+ \frac{1}{2f}\left(\delta^I_J df - h_{JK} h^{IL} \partial_{L}f \  \theta^K + \delta^I_K \partial_{J}f \ \theta^K \right),
\end{eqnarray}
with $df:=\partial_{I}f \ \theta^I$ and $h^{IJ}$ being the inverse of $h_{IJ}$. Also, the indices of $M_{IJ}$ in the r.h.s. have raised and lowered with $h_{IJ}$ and its inverse and the indices of the Levi-Civita symbol $\varepsilon^{ijk}$ are raised and lowered with $\delta_{ij}$.

The solution (\ref{Asol2}), obtained from linking the internal connection $A^i$ with the connection $\Omega^{I}{}_{J}$ defined by (\ref{defOmega}) and (\ref{defOmega2}), shows that the internal  connection $A^i$ is not expressed in terms of the spin connection $ \omega^{I}{}_{J}+ \frac{1}{2f}\left(\delta^I_J df - h_{JK}h^{IL} \partial_{L}f \  \theta^K + \delta^I_K \partial_{J}f \ \theta^K \right)$ for the case $T^I=0$ and $M_{IJ}=0$ in (\ref{defOmega}) and (\ref{defOmega2}) as one might naively think, but rather it is only expressed in terms of the spin connection $\omega^{I}{}_{J}$ defined by Eqs. (\ref{def3}) and (\ref{def4}) with $g_{IJ}=\eta_{IJ}$ and considered in SubSection 3.1. This allows us to conclude that to take the metric $\eta_{MN}$ or other conformally equal to $\eta_{MN}$ will lead us to the same theory: general relativiy.

\section{Concluding remarks}
We conclude by making some remarks and pointing out some consequences of the issue studied here and of our results, reported mainly in the section 3 of this paper:
\begin{enumerate}
\item
We have analyzed the relationship between the most general spacetime connection and the internal connection involved in Pleba\'nski's formulation of complex general relativity. Our approach involves, in addition to the reality conditions (\ref{reality1}) and (\ref{reality2}), the hypotheses 1 and 2 contained in Sect. 3 which amounts to use the Urbantke metric (\ref{urbmetric}) and Eqs. (\ref{def}) and (\ref{def2}). Once these ingredients are combined with Eqs. (\ref{pleb}) and (\ref{pleb2}) the result is that the internal connection $A^i$ is not expressed in terms of the full connection $\Gamma^I{}_J$ given in Eq. (\ref{conn}) or Eq. (\ref{connOmega}), but just in terms of $\omega^I{}_J$ defined by Eqs. (\ref{def3}) and (\ref{def4}) with $g_{IJ}=\eta_{IJ}$. 
\item
Closely related with the previous remark is the following fact. As should be clear to the reader by now, the emergence of general relativity from Pleba\'nski's equations of motion (\ref{einstein}), (\ref{pleb}), and (\ref{pleb2}) is not something direct nor should be taken for granted as it is usually stated. It arises because of the hypotheses assumed in Section 3, which are collected  in the previous item 1. If any of these hypotheses is changed the resulting theory might be something else, something different from general relativity.
\item
The results presented here are also useful for understanding the geometrical meaning of the connection
involved in Krasnov's modification of Pleba\'nski's action proposed some years ago \cite{krasnov,krasnov2,krasnov3}. This is so because, as in the Pleba\'nski formulation, in Krasnov's proposal the link between the internal and the spacetime connection has also to be set from extra hypotheses that do not come out solely from the handling of the equations of motion. This issue has been studied in detail within a perturbative treatment in Ref. \cite{merced} and, more recently, in a exact way in Ref. \cite{thesisdiego} using the previous results found in Ref. \cite{thesismeche}. All of these results will be reported soon somewhere else. For the moment it is enough to say that the results reported here for the Pleba\'nski formulation also apply for Krasnov's modification of it.
\item The strategy used in the Pleba\'nski formulation to recover general relativity can also be used in any other Pleba\'nski-like constrained BF theory based in any arbitrary Lie group.
\item Finally, the issue studied here is also relevant for the so-called pure spin-connection formulation of general relativity \cite{capo2} or any other theory based on (internal) gauge connections as Yang-Mills theory and the MacDowell-Mansouri  formulation of general relativity \cite{macdowell}, in which spacetime geometrical structures can be built up from internal ones.
\end{enumerate}

\begin{acknowledgements}
We thank Riccardo Capovilla  and Gerardo F. Torres del Castillo for useful discussions.  We also acknowledge
partial support from CONACyT grant 167477-F. M. Vel\'azquez's postdoctoral position in the {\it Centre de Physique Th\'eorique} is funded through a fellowship from CONACyT of M\'exico, fellow number 172172.
\end{acknowledgements}

\appendix

\section{From Einstein to Pleba\'nski equations of motion}\label{appendixA}
\setcounter{section}{1}
This appendix is based on Ref. \cite{thesismeche} and contains a pedagogical deduction of Pleba\'nski's equations for complex general relativity in the framework of the $so(3)$ formalism from Einstein's equations in the first-order formalism.

In the theoretical framework of the first-order formalism of general
relativity the {\it kinematical equations}\/ are the {\it first Cartan structural equations}
\begin{eqnarray}\label{fc}
d \theta^I + \omega^I{}_J \wedge \theta^J = 0,
\end{eqnarray}
the {\it second Cartan structural equations}
\begin{eqnarray}\label{sc}
R^I{}_J := d \omega^I{}_J + \omega^I{}_K \wedge \omega^K{}_J,
\end{eqnarray}
the {\it first Bianchi identities}
\begin{eqnarray}\label{fbi}
R^I{}_J \wedge \theta^J =0,
\end{eqnarray}
and the {\it second Bianchi identities}
\begin{eqnarray}\label{sbi}
d R^I{}_J + \omega^I{}_K \wedge R^K{}_J - \omega^K{}_J
\wedge R^I{}_K =0,
\end{eqnarray}
where it has been assumed that the connection $\omega$ is torsionless. In the
preceding equations, $\{ \theta^I \}$ is the dual basis of the basis formed by tangent vectors $\{e_I \}$.

The kinematical equations (\ref{fc})-(\ref{sbi}) define a topological field theory \cite{topo,topo2,topo3} in the sense that it has no local degrees of freedom. The topological property of Eqs. (\ref{fc})-(\ref{sbi}) is broken down by adding additional equations that make the full theory a theory of the
gravitational field. The {\it dynamical equations}\/ are precisely those given by Einstein's
equations
\begin{eqnarray}\label{gr}
\ast R^I{}_J \wedge \theta^J = 0,
\end{eqnarray}
where $\ast R_{IJ}= \frac12 \varepsilon_{IJ}{}^{KL} R_{KL}$ is the dual of
$R_{IJ}$. The indices $I,J,K \ldots$ are rised and lowered with the Minkowski metric $(\eta_{IJ})=(-1,1,1,1)$ and $\omega^I{}_J$ is assumed to be compatible with $\eta_{IJ}$, i.e, $d \eta_{IJ} - \omega^K{}_I \eta_{KJ} - \omega^K{}_J \eta_{IK}=0$.  Here $\varepsilon_{0123}=+1$ (and so $\varepsilon^{0123}=-1$).

{\it Self-dual substructure}. It is remarkable that real,
Lorentzian general relativity can be rewritten in terms of complex variables
using just self-dual variables essentially \cite{plebanski}. In what follows, it
is explained how to do that following, basically, Pleba\'nski's procedure.

In fact, eqs. (\ref{fc}) can alternatively be rewritten in terms of complex
variables as
\begin{eqnarray}\label{em1}
D \Sigma^i := d \Sigma^i + \varepsilon^i{}_{jk} A^j \wedge \Sigma^k =0, \quad
i,j,k=1,2,3,
\end{eqnarray}
where $A^i = \Gamma^i + \rmi \omega^{0i}$ is the self-dual connection 1-form
whose real part is $\Gamma^i = -\frac12 \varepsilon^i{}_{jk} \omega^{jk}$ and
$\Sigma^i$ is a set of three complex-valued 2-forms given by
\begin{eqnarray}\label{twoforms}
\Sigma^i = \theta^0 \wedge \theta^i + \frac{\rmi}{2} \varepsilon^i{}_{jk} \theta^j \wedge \theta^k
\, .
\end{eqnarray}
From their definition, $\Sigma^i$ satisfy\footnote{Note that the reality conditions are $\Sigma^i \wedge {{\overline \Sigma}}^j=0$ and $\Sigma^i \wedge \Sigma_i + {\overline \Sigma}^i \wedge {\overline \Sigma}_i=0$.}
\begin{eqnarray}\label{em2}
\Sigma^i \wedge \Sigma^j = \frac{\delta^{ij}}{3} \left ( \Sigma^k \wedge
\Sigma_k \right ),
\end{eqnarray}
with
\begin{eqnarray}
\Sigma^k \wedge \Sigma_k  = 6 \rmi \theta^0 \wedge \theta^1 \wedge \theta^2 \wedge \theta^3 \neq 0.
\end{eqnarray}

On the other hand, the first Bianchi identities (\ref{fbi}) and Einstein's
equations (\ref{gr}) imply that there are only 10 independent components in the curvature 2-forms $R^I{}_J$ \cite{Wheeler}, given by $E^i{}_j$ and $H^i{}_j$, where $(E^i{}_j)$ and $(H^i{}_j)$ are trace-free, symmetric, real matrices. More precisely, the components of the curvature 2-forms $R^I{}_J$ with respect to the orthonormal frame, $R^I{}_J = \frac12 R^I{}_{JKL} \theta^K \wedge \theta^L$, can be written in terms of these 10 variables as \begin{eqnarray}\label{clave2}
 R^{0i}{}_{0j} &=&  E^i{}_j, \nonumber\\
 R^{0i}{}_{jk} &=& \varepsilon_{jk}{}^l H^i{}_l, \nonumber\\
 R^{jk}{}_{0i} &=& - \varepsilon^{jk}{}_l H^l{}_i, \nonumber\\
 R^{ij}{}_{kl} &=& \varepsilon^{ij}{}_m \varepsilon_{kl}{}^n E^m{}_n.
\end{eqnarray}
The relevance of eqs. (\ref{clave2}) is that they allow us to relate the
self-dual connection $A^i$ through its curvature with the 2-forms $\Sigma^i$. These relations can
be obtained in the following way. Using the fact that the curvature $F^i$ of the
self-dual connection $A^i$ is the self-dual part of the curvature of $\omega^I{}_J$, $- \frac12 \varepsilon^i{}_{jk} R^{jk} + \rmi R^{0i}$, we have
\begin{eqnarray}
F^i &=&  - \frac12 \varepsilon^i{}_{jk} R^{jk} + \rmi R^{0i} \nonumber\\
&=&  - \frac14 \varepsilon^i{}_{jk} R^{jk}{}_{IJ} \theta^I \wedge \theta^J +
\frac{\rmi}{2} R^{0i}{}_{IJ} \theta^I \wedge \theta^J \nonumber\\
&=&  - \frac12 \varepsilon^i{}_{jk} R^{jk}{}_{0l} \theta^0 \wedge \theta^l - \frac{1}{4}
\varepsilon^i{}_{jk} R^{jk}{}_{lm} \theta^l \wedge \theta^m \nonumber\\
& & + \rmi R^{0i}{}_{0l} \theta^0 \wedge \theta^l + \frac{\rmi}{2} R^{0i}{}_{lm} \theta^l \wedge \theta^m,
\end{eqnarray}
where in the second and third equalities $R^I{}_J = \frac12 R^I{}_{JKL} \theta^K
\wedge \theta^L$ was used. By inserting eqs. (\ref{clave2}) into the RHS of last
equality and using the definition of $\Sigma^i$ given in eq. (\ref{twoforms}) it is obtained 
\begin{eqnarray}\label{em3}
F^i = C^i{}_j \Sigma^j,
\end{eqnarray}
where $C^i{}_j = H^i{}_j + \rmi E^i{}_j $ is a symmetric complex matrix which
is also trace-free
\begin{eqnarray}\label{em4}
C^i{}_i=0,
\end{eqnarray}
because of $E^i{}_i=0$ and $H^i{}_i=0$.

Notice that eqs. (\ref{em1}), (\ref{em2}), (\ref{em3}), and (\ref{em4}) involve just the self-dual connection 1-form $A^i$, the 2-forms $\Sigma^i$, and the scalar fields $C_{ij}$. All of these variables are complex-valued. These equations constitute one of the Einsteinian substructures in Pleba\'nski's terminology, namely, the self-dual substructure.Therefore, instead of using the tetrad field $\theta^I$ and the spin connection 1-form $\omega^I{}_J$ to describe (real and Lorentzian) general relativity, in Plebanski's formulation, the fundamental variables are $A^i$, $\Sigma^i$, and $C_{ij}$. Note also, that the field variables $C_{ij}$ are {\it dynamical variables} in the sense that obey a differential equation which is explicitly exhibited only after the use of the Bianchi identities for the curvature of the self-dual connection themselves, $D F^i=0$, and the equations of motion
(\ref{em1}) and (\ref{em3}). In fact, using eqs. (\ref{em3}) one gets $D F^i=D C^i{}_j \wedge \Sigma^j + C^i{}_j D \Sigma^j$ and so 
\begin{eqnarray}
D C^i{}_j \wedge \Sigma^j=0,
\end{eqnarray}
because of the Bianchi identities $D F^i=0$ and eqs. (\ref{em1}).

Pleba\'nski gave an action principle \cite{plebanski} that allows us to obtain Eqs. (\ref{em1}), (\ref{em2}), (\ref{em3}), and (\ref{em4}).
Nevertheless, people usually employ an equivalent and more economical action principle in which the equation (A.13) is solved from the very beginning; this action principle is given by
\begin{eqnarray}\label{eq:accpleba}
S[A^i, \Sigma^i, C_{ij}]=\int _{\mathcal{M}^4} \left[ \Sigma_i \wedge F^i[A] -\frac{1}{2} C_{ij} \Sigma^i \wedge \Sigma^j   \right].
\end{eqnarray}
Both action principles are equivalent. The variation of action (\ref{eq:accpleba}) with respect to independent fields leads to the equations of motion (\ref{em1}), (\ref{em2}), and (\ref{em3}) [or equivalently  (\ref{einstein}), (\ref{pleb}), and (\ref{pleb2})]
\begin{eqnarray}
\delta \Sigma^i: &  F^i = C^i{}_j \Sigma^j , \quad C^i{}_i=0, &\quad(3\times 6=18 \quad\textnormal{equations}),\label{eq:1pleban}\\
\delta A^i:         & d \Sigma^i  +  \varepsilon^i{}_{jk} A^j \wedge \Sigma^k =0,  &\quad(3\times 4=12 \quad\textnormal{equations}),\label{eq:3pleban}\\
\delta C_{ij}: & \Sigma^i \wedge \Sigma^j - \frac13 \delta^{ij} \Sigma^k \wedge \Sigma_k=0,  &\quad (5 \quad\textnormal{equations}), \label{eq:4pleban}
 \end{eqnarray}
where it has been assumed that $C_{ij}$ satisfies (\ref{em4}) from the very beginning. Note that there are $3\times 4=12$ variables in $A^i$, $3\times 6=18$ variables in $\Sigma^i$ and $5$ variables in the traceless, symmetric matrix $C_{ij}$. Therefore, there are 35 variables and 35 equations involved, so the system of equations is closed.

\section{The solution for the internal connection}\label{appendixB}

In this appendix we show explicitly that for a generic $\Sigma^i = \frac12 \Sigma^i{}_{IJ} \theta^I \wedge \theta^J$, the solution of the system of linear equations (\ref{pleb}) for the unknowns $A^i$ is (\ref{connex}) when the linear system is non-degenerate. In order to do this, the equations (\ref{pleb}) can be rewritten as
\begin{eqnarray}\label{eq:AppBequiv}
d\widetilde\Sigma^{iL} + \varepsilon^{i}{}_{jk} A^j{}_N \widetilde \Sigma^{kNL}=0,
\end{eqnarray}
where $A^i=A^i{}_J \theta^J$, $\widetilde\Sigma^{iKL} := \frac12 \Sigma^i{}_{IJ} {\widetilde\eta}^{IJKL}$, $d\Sigma^i = \frac{1}{3!} d\Sigma^i{}_{IJK} \theta^I \wedge \theta^J \wedge \theta^K$, $d\widetilde\Sigma^{iL} := \frac{1}{3!} d\Sigma^i{}_{IJK} {\widetilde\eta}^{IJKL}$ with ${\widetilde\eta}^{IJKL}$ being the totally antisymmetric Levi-Civita symbol (with ${\widetilde\eta}^{0123}=+1$). Equivalently, we can rewrite the 12 linear equations (\ref{eq:AppBequiv}) for the 12 unknowns $A^i{}_N$ in the form
\begin{eqnarray}\label{eq:AppBmtxform}
M {\bf x}={\bf b},
\end{eqnarray}
where ${\bf x}$ is a column vector with 12 entries $x_1:=A^1{}_0,x_2:=A^1{}_1,\ldots,x_{11}:=A^3{}_2,x_{12}:=A^3{}_3$, ${\bf b}$ is a column vector with 12 entries $b_1:=-d\widetilde\Sigma^{10},b_2:=-d\widetilde\Sigma^{11},\ldots,b_{11}:=-d\widetilde\Sigma^{32},b_{12}:=-d\widetilde\Sigma^{33}$ and $M$ is a $12\times12$ matrix given by
\begin{eqnarray}
M=\left( \begin{array}{rrr}
0_{4\times4} & M^{3} & -M^{2} \\
-M^{3} & 0_{4\times4} & M^{1} \\
M^{2} & -M^{1} & 0_{4\times4}
\end{array} \right),
\end{eqnarray}
with
\begin{eqnarray}
M^i=\left( \begin{array}{rrrr}
0 & -\Sigma^i{}_{23} & -\Sigma^i{}_{31} & -\Sigma^i{}_{12} \\
\Sigma^i{}_{23} & 0 & -\Sigma^i{}_{03} & \Sigma^i{}_{02} \\
\Sigma^i{}_{31} & \Sigma^i{}_{03} & 0 & -\Sigma^i{}_{01} \\
\Sigma^i{}_{12} & -\Sigma^i{}_{02} & \Sigma^i{}_{01} & 0 
\end{array} \right ), \quad i=1,2,3.
\end{eqnarray}
Now, we treat only the case in which $M$ is non-degenerate. With this in mind, the unique solution of (\ref{eq:AppBmtxform}) is
\begin{eqnarray}\label{eq:AppBmtxsol}
{\bf x}=M^{-1}{\bf b},
\end{eqnarray}
where 
\begin{eqnarray}
M^{-1}=\left( \begin{array}{ccc}
\Psi^{11} & \Psi^{12} & \Psi^{13} \\
\Psi^{21} & \Psi^{22} & \Psi^{23} \\
\Psi^{31} & \Psi^{32} & \Psi^{33} 
\end{array} \right), 
\end{eqnarray}
is the inverse matrix of $M$ with $4\times4$ sub-matrices $\Psi^{ij}$ whose elements are 
\begin{eqnarray}
\Psi^{ij}{}_{MN} = -8 \left( \frac{m^{ij} k_{MN}}{\det{(m^{ij})}} +\frac{1}{4} \varepsilon^{ijk} (m^{-1})_{kl} \Sigma^l{}_{MN} \right ), \label{appconnexPsi}
\end{eqnarray}
where
\begin{eqnarray}
{k}_{MN} := -\frac{1}{12} \varepsilon_{ijk} \Sigma^i{}_{MI} \Sigma^j{}_{JK} \Sigma^k{}_{LN} \ {\widetilde \eta}^{IJKL},  \label{appurbmetric}
\end{eqnarray}
\begin{eqnarray}
m^{ij}:=\frac12 \Sigma^i{}_{IJ} \Sigma^j{}_{KL} \ {\widetilde \eta}^{IJKL}. \label{appmpleb}
\end{eqnarray}
Here, ${k}_{MN}$ is the Urbantke metric with respect to an arbitrary basis of 1-forms $\{\theta^1, \theta^2, \theta^3, \theta^4\}$ and $(m^{-1})_{ij}$ denotes the inverse of $m^{ij}$ ($(m^{-1})_{ik} m^{kj} = \delta^j_i$). The lower-case latin indices in the 2-forms are raised and lowered with Kronecker delta $ \delta_{ij}$. In components, the solution (\ref{eq:AppBmtxsol}) acquires the form
\begin{eqnarray}
A^i{}_{T} = - \frac{1}{3!} \Psi^i{}_{jTR} d \Sigma^j{}_{IJK} {\widetilde\eta}^{IJKR}.\label{appconnex}
\end{eqnarray}
Notice that the solution   (\ref{appconnex}) is generic in the sense that the explicit form for the $\Sigma$'s that solves  (\ref{pleb2}) was not used and it holds if $\det{M} =16 \det{(k_{MN})} =(\frac{1}{16} \det{(m^{ij})})^2  \neq 0$. Since the systems of equations (\ref{pleb}), (\ref{eq:AppBequiv}) and (\ref{eq:AppBmtxsol}) are equivalent in the non-degenerate case, the formula (\ref{appconnex}) that solves  (\ref{eq:AppBmtxsol})  also solves (\ref{pleb}) and (\ref{eq:AppBequiv}).


\begin{thebibliography}{99}

\bibitem{trautman}
A. Trautman,
``Fibre bundles associated with space-time,"
\href{http://dx.doi.org/10.1016/0034-4877(70)90003-0}{\it Rep. Math. Phys.} (Torun) {\bf 1}, 29--62 (1970).

\bibitem{civita}
T. Levi-Civita,
\textit{The Absolute Differential Calculus (Calculus of Tensors)}
(Dover Publications, Inc., Mineola, New York, 1977).

\bibitem{plebanski}
J.F. Pleba\'nski,
``On the separation of Einsteinian substructures,"
\href{http://dx.doi.org/10.1063/1.523215}{\JMP} {\bf 18}, 2511--2520 (1977).

\bibitem{capo} 
R. Capovilla, J. Dell, T. Jacobson, and L. Mason,
``Self-dual 2-forms and gravity,"
\href{http://dx.doi.org/10.1088/0264-9381/8/1/009}{\CQG} {\bf 8}, 41--57 (1991).

\bibitem{bengtsson}
I. Bengtsson,
``2-form geometry and the 't Hooft-Plebanski action,"
\href{http://dx.doi.org/10.1088/0264-9381/12/7/004}{\CQG} {\bf 12}, 1581--1950 (1995).

\bibitem{krasnov4}
K. Krasnov,
``Pleba\'nski formulation of general relativity: a practical introduction," \href{http://dx.doi.org/10.1007/s10714-010-1061-x}{\it Gen. Rel. Grav.} {\bf 43}, 1--15 (2011), \href{http://arxiv.org/abs/0904.0423}{arXiv:0904.0423}.

\bibitem{krasnov5}
K. Krasnov,
``Effective metric Lagrangians from an underlying theory with two propagating degrees of freedom,"
\href{http://dx.doi.org/10.1103/PhysRevD.81.084026}{\PR D} {\bf 81}, 084026, 40~pages (2010),
\href{http://arxiv.org/abs/0911.4903}{arXiv:0911.4903}.

\bibitem{cartan}
E. Cartan,
\textit{Riemannian Geometry in an Orthogonal Frame}
(World Scientific, Singapore, 2001).

\bibitem{thesisdiego}
D. Gonz\'alez,
\textit{Gravedad de 2-formas},
M.Sc. thesis, Cinvestav, Mexico, (2011).

\bibitem{tdiego}
D.G. Gonz\'alez Vallejo,
\textit{Gravedad de 2-formas}
(Editorial Acad\'emica Espa\~nola, Saarbr\"ucken, 2012).

\bibitem{torres}
G.F. Torres del Castillo,
\textit{Differentiable Manifolds. A Theoretical Physics Approach}
(\href{http://dx.doi.org/10.1007/978-0-8176-8271-2}{Birkh\"auser}, Boston, 2012).

\bibitem{urbantke}
H. Urbantke,
``On integrability properties of SU (2) Yang-Mills fields. I. Infinitesimal part,"
\href{http://dx.doi.org/10.1063/1.526402}{\JMP} {\bf 25}, 2321--2324 (1984).

\bibitem{krasnov}
K. Krasnov,
``Renormalizable Non-metric quantum gravity?,"
(2007), \href{http://arxiv.org/abs/hep-th/0611182}{arXiv:hep-th/0611182}

\bibitem{krasnov2}
K. Krasnov,
``Non-metric gravity: A status report,"
\href{http://dx.doi.org/10.1142/S021773230702590X}{\it Mod. Phys. Lett. A}  {\bf 22}, 3013--3026 (2007),
\href{http://arxiv.org/abs/0711.0697}{arXiv:0711.0697}.

\bibitem{krasnov3}
K. Krasnov,
``Deformations of the constraint algebra of Ashtekar's Hamiltonian formulation of general relativity,"
\href{http://dx.doi.org/10.1103/PhysRevLett.100.081102}{\PRL} {\bf 100}, 081102, 4~pages (2008).

\bibitem{merced}
R. Capovilla, M. Montesinos, and M. Vel\'azquez,
``Minimally modified self-dual 2-forms gravity,"
\href{http://dx.doi.org/10.1088/0264-9381/27/14/145011}{\CQG} {\bf 27}, 145011, 6~pages ( 2010),
\href{http://arxiv.org/abs/1004.3324}{arXiv:1004.3324}.

\bibitem{thesismeche} 
M. Vel\'azquez,
\textit{BF gravity, matter couplings, and related theories},
Ph.D. thesis, Cinvestav, Mexico,  (2011).

\bibitem{capo2}
R. Capovilla, J. Dell, and T. Jacobson,
``A pure spin-connection formulation of gravity,"
\href{http://dx.doi.org/10.1088/0264-9381/8/1/010}{\CQG} {\bf 8}, 59--73 (1991).

\bibitem{macdowell}
S. MacDowell and F. Mansouri,
``Unified geometric theory of gravity and supergravity,"
\href{http://dx.doi.org/10.1103/PhysRevLett.38.739}{\PRL} {\bf 38}, 739--742 (1977), 
Erratum,
\href{http://dx.doi.org/10.1103/PhysRevLett.38.1376}{\PRL} {\bf 38}, 1376 (1977).

\bibitem{topo} 
V. Cuesta and M. Montesinos,
``Cartan's equations define a topological field theory of the BF type,"
\href{http://dx.doi.org/10.1103/PhysRevD.76.104004}{\PR D} {\bf 76}, 104004, 6~pages (2007).

\bibitem{topo2} 
V. Cuesta, M. Montesinos, M. Vel\'azquez, and J.D. Vergara,
``Topological field theories in n-dimensional spacetimes and Cartan's equations,"
\href{http://dx.doi.org/10.1103/PhysRevD.78.064046}{\PR D} {\bf 78}, 064046, 10~pages (2008),
\href{http://arxiv.org/abs/0809.2741}{arXiv:0809.2741}.

\bibitem{topo3} 
L. Liu, M. Montesinos, and A. Perez,
``Topological limit of gravity admitting an SU(2) connection formulation,"
\href{http://dx.doi.org/10.1103/PhysRevD.81.064033}{\PR D} {\bf 81}, 064033, 9~pages (2010),
\href{http://arxiv.org/abs/0906.4524}{arXiv:0906.4524}.

\bibitem{Wheeler} 
C.W. Misner, K.S. Thorne, and J.A. Wheeler,
\textit{Gravitation}
(W. H. Freeman and Co., San Francisco, 1973).

\end{thebibliography}
\end{document}